\documentclass[aps,prd,twocolumn,superscriptaddress,nofootinbib,showpacs,showkeys]{revtex4-1}
\usepackage{graphicx,color}
\usepackage{rotating}
\usepackage{amssymb}
\usepackage{amsfonts}
\usepackage{amsmath}
\usepackage{lipsum}

\newcommand{\be}{\begin{equation}}
\newcommand{\ee}{\end{equation}}
\newcommand{\bea}{\begin{eqnarray}}
\newcommand{\eea}{\end{eqnarray}}

\date{\today}

\begin{document}

\title{Nonextensive Black Hole Entropy and Quantum Gravity Effects at the Last Stages of Evaporation}

\author{Ana Alonso-Serrano}
\email[]{ana.alonso.serrano@aei.mpg.de}
\affiliation{Max-Planck-Institut f\"ur Gravitationsphysik  (Albert-Einstein-Institut), Am M\"{u}hlenberg 1, D-14476 Potsdam, Germany}
\author{Mariusz P. D\c{a}browski}
\email[]{mariusz.dabrowski@usz.edu.pl}
\affiliation{Institute of Physics, University of Szczecin, Wielkopolska 15, 70-451 Szczecin, Poland}
\affiliation{National Centre for Nuclear Research, Andrzeja So{\l}tana 7, 05-400 Otwock, Poland}
\affiliation{Copernicus Center for Interdisciplinary Studies, S{\l}awkowska 17, 31-016 Krak\'ow, Poland}
\author{Hussain Gohar}
\email[]{hussain.gohar@sns.nust.edu.pk} 
\affiliation{Department of Physics, School of Natural Sciences, National University of Sciences and Technology, H-12, 44000 Islamabad, Pakistan}
\begin{abstract}
We analyze the Generalized Uncertainty Principle (GUP) impact onto the nonextensive black hole thermodynamics by using R\'enyi entropy. We show that when introducing GUP effects, both R\'enyi entropy and temperature associated with black holes have finite values at the end of the evaporation process. We also study the sparsity of the radiation, associated with R\'enyi temperature, and compare it with the sparsity of Hawking radiation. Finally, we investigate GUP modifications to the sparsity of the radiation when GUP effects are introduced into R\'enyi temperature.

\end{abstract} 
\maketitle
\section{Introduction}
The thermodynamical study of black holes goes back to the seminal works of Bekenstein and Hawking \cite{be,hr1}, where the derivation of the laws of thermodynamics of black holes was performed once quantum field theory effects were introduced \cite{hr2a, hr2}. Those effects allow the black hole to thermodynamically interact with the environment and can result in its evaporation by means of the emission of the Hawking radiation \cite{hr3}. Since then, a large number of studies have been developed to fully understand the Hawking radiation of black holes \cite{hr4,hr5,hr6,hr7,hr8,hr9,hr10,hr10a,hr13,hr14,hr15,hr16}. But still, there is no conclusive picture of the black hole evaporation process that is physically consistent and complete \cite{hr11,hr12}. Moreover, the lack of a final theory of quantum gravity prevents a full understanding of the nature of this process.

Heuristically, it has been shown that the temperature of a black hole can be deduced by using the uncertainty relation $\Delta p\Delta x\approx \hbar$, where $x$ and $p$ are the position and momentum of a particle and $\hbar$ is the reduced Planck constant respectively \cite{gup1,GUPReview}. Due to the effects coming from quantum field theory in the vicinity of the horizon, one can consider uncertainty in the position of a particle \cite{gup1}. Considering the minimum position uncertainty near the event horizon of the Schwarzschild black hole as $\Delta x=2l_p=4GM/c^2$, where $l_p$ is the Planck length, the energy uncertainty can be written as
\be
\Delta pc\approx \frac{\hbar c^3}{4 GM} \label{energy},
\ee 
where $M$ is the mass of the black hole, $G$, $c$  and $k_B$ are the speed of light, the Newton's gravitational constant, and the Boltzmann's constant, respectively. By introducing a calibration factor of $2\pi$, the Hawking temperature $T_{bh}$ can be expressed as
\be
T_{bh}=T_p\left(\frac{m_p}{8\pi M}\right) = \frac{c^3 \hbar}{8\pi G k_B M}, 
\label{tbh} 
\ee
where $T_p= m_p c^2/k_B$ and $m^2_p=\hbar c/G$ are the Planck's temperature and Planck's mass, respectively. By using the first law of thermodynamics, $c^2dM=T_{bh}dS_{bh}$, the entropy of the Schwarzschild black hole, $S_{bh}$, can be derived as
\be
S_{bh}=4\pi k_B\left(\frac{M}{m_p}\right)^2 = \frac{4\pi k_BGM^2}{\hbar c} . 
\label{sbh}
\ee
Note that the entropy $S_{bh}$ approaches zero, and the temperature $T_{bh}$ blows up to infinity when $M$ goes to zero during the Hawking evaporation process. These quantities correspond to a black hole that completely evaporates due to the emission of Hawking radiation.

During the final stages of the Hawking evaporation, the semiclassical approach is expected to break down, due to the dominance of quantum gravity effects. Although there exist very different proposals \cite{qg1,qg2,qg3,qg4,qg5,qg6,qg7,qg8}, there is not yet a satisfactory theory of quantum gravity that allows us to fully understand that regime. One way to study the quantum gravity effects near those scales is to consider the phenomenological effects of an underlying theory of quantum gravity \cite{qg1,qg2,qg3,qg4,qg5}. One approach, that has the advantage of being enough general to be consistent with several theories \cite{qg6,qg7},  is given by the Generalized Uncertainty Principle (GUP) \cite{gup2,gup3,gup4,gup5,gup6,gup7,gup8}. Within this framework, the entropy of a black hole at its last stages of evaporation is modified. It is worth noticing that in this approach there are two types of possible modifications: one comes from canonical corrections \cite{correction1} and the other one from microcanonical corrections \cite{correction2}. The canonical corrections are related to thermal fluctuations on the horizon and it results in an increment of entropy. The microcanonical corrections refer to a quantum modification in counting microstates while keeping the horizon area fixed. It reduces the entropy as a consequence of the reduction of the uncertainty in the underlying microstates.
 
One of the important features of the Hawking radiation, which differentiates it from the black body radiation, is its extreme sparsity during the black hole evaporation process \cite{sparse1,sparse2,sparse5,ana1,ana2,ong, greybody,bibhas,sparse6,sparse7,sparse8}. Sparsity is defined by the average time between the emission of successive Hawking quanta over the timescales set by the energies of the emitted quanta. It is shown that the Hawking radiation is sparse during the whole evaporation of a black hole \cite{sparse1}. However, it has also been shown that, when phenomenological quantum gravity effects (expressed by GUP) are included, the sparsity diminishes at the final stages of evaporation,  allowing for a final burst of emission (accelerating the evaporation) \cite{ana1,ana2}.

 Classical thermodynamics of macroscopic objects as derived from statistical physics is very well understood. In this formalism, the macroscopic properties of a system can be obtained from the microscopic description of the system. Usually, it is assumed that long-range interactions are negligible. This means that the size of the system under study is larger than the range of the interaction between the elements of the system. By applying this condition, the local extensive parameters of the system are well defined and the entropy function of the system is obtained by using the additive Boltzmann-Gibbs formula. Hence, in the macroscopic limit, the classical thermodynamics can be recovered. However, for strongly gravitating systems, long-range type interactions cannot be ignored. For example, for strong gravitational fields, like for the case of black holes, the usual local extensive parameters are not the most appropriate to work with \cite{Tsallis2009,Tsallis2020}. Therefore, Boltzmann-Gibb's definition of the additive entropy may not be a suitable choice for strongly gravitating systems, where the properties of the long-range interactions are taken into account. 
	
In this way, it has been shown \cite{davies} that Bekenstein entropy is a nonextensive quantity and some studies have been developed to understand black hole entropy and characteristic features of evaporation in the light of nonextensive thermodynamics \cite{stability1,chinernew,stability1a,stability1b,biro,reny6,reny7,reny4}. The natural entropies associated with nonextensive thermodynamics are R\'enyi entropy \cite{reny2,reny3} and Tsallis entropy \cite{tsallisbook,Tsallis2009,reny1}. Both entropies are related to each other and provide different generalizations to Boltzmann-Gibbs entropy. 

Generally, in nonextensive thermodynamics, the nonadditive entropy composition rules are not compatible with the requirement of thermal equilibrium and they do not satisfy the zeroth law of thermodynamics \cite{reny7}. To solve these problems, it was found \cite{reny7} that by using a simple transformation, the nonadditive entropy composition rule of a given system was mapped into an additive one using Abe formula \cite{reny8}. This method provides a well-defined entropy function for the system that satisfies the zeroth law and the natural requirement for the thermal equilibrium.

Furthermore, although standard black hole thermodynamics described by Hawking temperature and Bekenstein entropy, proved to be a very successful theory, it had also some problems that need to be solved. For instance, for large mass black holes in a very large bath of thermal radiation, black holes are unstable. Since the stability corresponds to the additive entropy function of the system, while the Bekenstein entropy is nonadditive and it has negative specific heat, then it is difficult to maintain black hole stability. It has been shown \cite{chinernew} that the isolated rotating black holes are stable within the framework of R\'enyi statistics. Besides, black holes surrounded by a thermal heat bath are in stable equilibrium with the surroundings at a fixed temperature. This is unlike the case of the additive Boltzmann-Gibbs statistics, where black holes are unstable with the surrounding heat bath.

In this paper, we focus on analyzing the effects of GUP onto the R\'enyi entropy and the R\'enyi temperature of black holes, and the sparsity of  R\'enyi radiation. Note that we use the term,  R\'enyi radiation for the evaporation of a black hole within the framework of nonextensive thermodynamics, that is, associated with R\'enyi temperature, to differentiate it from Hawking radiation.

The paper is organized as follows. In Section \ref{GUP}, we review the GUP modifications to Hawking temperature and Bekenstein entropy. In Section \ref{nonext}, we introduce R\'enyi  entropy and the corresponding R\'enyi temperature. Then, we study GUP modifications related to these quantities. In Section \ref{sparsity}, we introduce the sparsity parameter and analyze the sparsity of the radiation and its modification by GUP effects. In Section \ref{conclusion}, we present a discussion of the results.

\section{Generalized Uncertainty Principle Review}
\label{GUP}

One way to deal with phenomenological aspects of quantum gravity is by considering the effects coming from the existence of a minimum length. These effects are encoded in what was called the Generalized Uncertainty Principle \cite{gup2,gup3,gup4,gup5,gup6,gup7,gup8} and are predicted from several theories and developments of quantum gravity \cite{qg1,qg2,qg3,qg4,qg5,qg6,qg7,qg8}, providing a considerably general approach to low-energy quantum gravity effects. In any case, it is constrained to that regime and cannot provide a full theory. It has been proposed that the Heisenberg Uncertainty Principle is modified when including gravity into the game, due to the appearance of a minimum length at the Planck scale in some quantum gravity approaches. This Generalized Uncertainty Principle (GUP) \mbox{reads \cite{gup1,GUPReview,gup2,gup3,gup4,gup5,gup6}}
\be
\Delta x \Delta p = \hbar \left[1 + \alpha_0\frac{l_{p}^2}{\hbar^2} (\Delta p )^2 \right],
\label{GUPb}
\ee
where $\alpha_0$ is a dimensionless constant that is predicted to be order unity. The prediction is merely theoretical, due to the extension of quantum gravity effects, and, although there exist several observational and experimental studies placing bounds on its value \cite{alpha1,alpha2,alpha3,alpha3,alpha4,alpha5,alpha6, alpha7, alpha8, alpha9}, they are still far to provide realistic and effective constraints.

GUP modifies the Hawking temperature as  \cite{gup1}
\bea
T_{gup} = \frac{4T_{bh}}{ \left[2+\sqrt{4-\alpha_0\frac{ m_p^2}{M^2}}\right]}, \label{tgup}
\label{tgup1}
\eea
where $T_{bh}$ is the standard Hawking temperature that is consistently recovered in the limit $\alpha_0 \to 0$. The sign of the dimensionless parameter $\alpha_0$ plays a very important role here. For $\alpha_0>0$, the temperature $T_{bh}$ reaches a finite value in a finite time when a black hole mass approaches some critical mass $M_c$ during the Hawking evaporation process. On the other hand, for $\alpha_0<0$, the temperature has still finite value while the mass of the black hole approaches zero and yields infinite lifetime \cite{onga}.  It means that for positive values of $\alpha_0$; the evaporation process stops at $M_c=(\sqrt{\alpha_0}m_p)/2$, and the black hole does not evaporate completely. Therefore, the final state of the black hole is a remnant of order of Planck mass, having finite temperature $T_c=T_p/ (2\pi\sqrt{\alpha_0})$. On the other hand, for negative values of $\alpha_0$; the final stage of evaporation would be a so-called ``zero mass remnant'' (due to its asymptotical limit) \cite{onga}.  Similarly, there have been other studies regarding negative GUP parameter $\alpha_0$ \cite{ongc,ongb}.  

The modification of Hawking temperature gives rise to a GUP corrected entropy, $S_{gup}$, that can be written as
\bea 
&&S_{gup}=\frac{S_{bh}}{4} \left[2+\sqrt{4-\alpha_0\frac{ m_p^2}{M^2}}\right] \nonumber\\ 
&&-k_B\pi \frac{\alpha_0}{2}\ln \left[\frac{M}{M_0}  \left(2+\sqrt{4-\alpha_0\frac{ m_p^2}{M^2}}\right)\right] , 
\label{sgup}
\eea  
where $M_0$ is an integration constant with mass units. It can be seen that in the limit $\alpha_0 \to 0$ Bekenstein entropy is recovered, as expected. Whereas Bekenstein entropy $S_{bh}$ goes to zero when $M$ approaches zero, GUP modified entropy $S_{gup}$ has a finite value at $M_c=(\sqrt{\alpha_0}m_p)/2$ for $\alpha_0>0$. This is a consequence of the existence of a minimum length, that gives rise to the appearance of a remnant with finite entropy when the mass of the black hole approaches  $M_c$. We can easily find the entropy of the remnant at $M_c$ for $\alpha_0>0$, that reads
\be
S_c=\frac{\alpha_0\pi k_B}{2}\left(1-\ln\frac{\sqrt{\alpha_0}m_p}{M_0}\right) .
\ee
One possibility that was conjectured to solve the black hole information problem is a final state of evaporation as a remnant that can store information. However, this idea faces difficulties (see \cite{hr15} for a detailed discussion about different remnant scenarios and their relation to the black hole information problem). Let us remark that $S_{gup}$ for $\alpha_0<0$, although giving a positive correction to entropy (it is adding uncertainty due to the fluctuations), decreases faster at the latest stages of evaporation and it reaches zero at some finite mass $M$. Further, when $M$ approaches zero, the zero mass remnant has negative entropy. Because of that, we will not consider these corrections in our study because the final stage of evaporation and its properties are still not completely clear. It needs to be studied in detail in future work to check the viability of these corrections and their predictions. As a result, through the rest of the paper, we will focus on the well-understood case of $\alpha_0>0$.

\section{GUP and the Nonextensive Black Hole Thermodynamics} 
\label{nonext}

In Boltzmann-Gibbs thermodynamical description, the entropy is an additive quantity, which means that the entropy of the total isolated system is equal to the sum of the entropies of the two isolated subsystems i.e. that $S_{12}=S_1+S_2$, where $S_{12}$ is the entropy of the total system and $S_1$ and $S_2$ are the entropies of corresponding subsystems. 
As it was shown \cite{davies,stability1}, the additivity of entropy is not the case for the black hole thermodynamics since Bekenstein entropy is not an extensive parameter and fulfils the following nonadditive composition rule
\be
S_{12} = S_1 + S_2 + 2 \sqrt{S_1}\sqrt{S_2} .
\label{BekRule}
\ee
For black holes $S \propto M^2$, and when two black holes of masses $M_1$ and $M_2$ merge adiabatically, before the merger the sum of their entropies 
is $M_1^2/4+M_2^2/4$, while after the merger, the entropy jumps by a factor of $M_1M_2/2$ and is $(M_1+M_2)^2/4$ fulfilling the above mentioned composition rule (\ref{BekRule}). Note that this composition rule is also held by Tsallis entropy, providing the original motivation for the interpretation of Bekenstein entropy as Tsallis entropy \cite{stability1}.  In order to generalize the previous case, it can be introduced a nonadditive parameter $\lambda$ fulfilling the following rule
\be
S_{12} = S_1 + S_2 + \lambda S_1 S_2.
\label{RenRule}
\ee
Both composition rules are examples of the nonadditive entropy composition rule of Abe \cite{reny8} 
\be
H(S_{12}) = H(S_1) + H(S_2) + \lambda H(S_1) H(S_2),
\label{Hrule}
\ee
with $H(S)$ being a differentiable function of entropy that turns to be additive when considering a general logarithm of the form
\be
L(S) = \frac{1}{\lambda} \ln{\left[1 + \lambda H(S) \right]} ,
\label{LRenyi}
\ee
that fulfills 
\be 
L(S_{12}) = L(S_1) + L(S_2) .
\label{addL}
\ee
It has been shown that, in fact, funcion $L(S)$ corresponds to the definition of R\'enyi entropy and $H(S)$ can be identified with Tsallis entropy \cite{chinernew}.

In statistical terms, Tsallis entropy can be defined \mbox{as \cite{reny1,tsallisbook}} 
\be
S_T= k_B\left[\frac{1-\sum_{i=1}^Wp_i^q}{q-1}\right],
\label{STsallis}
\ee
for a set of $W$ discrete states, where $p_i$ (with $\sum_{i=1}^Wp_i =1$) is a probability distribution and $q \in \mathbb{R}$, $q\neq 1$, is a dimensionless nonextensivity parameter. In fact, the Tsallis entropy generalizes the Boltzmann-Gibbs statistics for  strongly coupled systems \cite{tsallisbook}, where the extensive nature of entropy does not work. On the other hand, the R\'enyi entropy $S_{R}$ \cite {reny2,reny3} is defined in the following way 
\be
S_{R}= k_B\left[\frac{\ln\sum_{i=1}^Wp_i^q}{1-q}\right],
\ee 
with $q\geq 0$ and $q\neq 1$, that can be written in terms of $S_T$
\be
S_{R}=\frac{k_B}{1-q}\ln[1+(1-q)\frac{S_T}{k_B}] 
\label{ST} ,
\ee
and it corresponds to the definition (\ref{LRenyi}), if we define $\lambda = 1-q$. 
Note that for $q\rightarrow 1$ or $\lambda \to 0$ (or $k_B \to \infty$, cf. \cite{Tsallis2009}), both $S_T$ and $S_R$ reduce to the standard Boltzmann-Gibbs entropy (Shannon entropy)
\be
S_{BG}=-k_B\left[\sum_{i=i}^Wp_i\ln p_i\right].
\ee
In fact, each of these entropies provides a family of $q$-entropies. The value of $q$ parameter determines the order of the entropy and it is crucial for its interpretation \cite{reny2,reny3,Tsallis2009}. For the case of R\'enyi entropy that we will focus on this paper, some particular values of $q$ are well know in the literature (as e.g. $q=0$ for Hartley or max-entropy and $q=2$ for critical or collision entropy) \cite{GazeauTsallis2019,Tsallis2020,Amigo2018,Duch2008,Wilk2000}. It is worth mentioning that for the values of $q \geq 3/2$, the standard mean value of the energy of the $q$-generalized canonical ensemble \cite{Tsallis2009} is infinite. For various complex systems, one finds many specific values of the $q$-index, which can be determined from the data \cite{Tsallis2009}. For example, for the so-called $q$-triplet which composes of $q_{sensitivity}$ that describes sensitivity to the initial conditions of the system, $q_{relaxation}$ that characterizes the dissipation of the nonlinear dynamical system, and $q_{correlation}$ that measures the degree of correlation in the system, one finds the values of $q$ in the range from $-0.6$ up to $3.8$. However, bearing in mind the requirement of the positivity of entropy for black holes in Eq.~(\ref{ST}), it is advisable to restrict oneself to $q<1$.  

Considering that the entropy of a black hole can be interpreted as a nonextensive entropy defined by Tsallis entropy, $S_T=\hat S_{bh}$ \cite{biro,stability1}, one can introduce  R\'enyi entropy associated to a black hole \cite{biro,stability1} as
\be
S_{R}=\frac{k_B}{\lambda}\text{ln}\left[1+\lambda \hat S_{bh}\right], 
\label{SR}
\ee 
 where $\hat S_{bh}= S_{bh}/k_B$ is a dimensionless entropy measured in bits.  Note that in analyzing R\'enyi entropy for black holes, due to previous reasoning, we will restrict our study to  R\'enyi entropies with $q<1  (\lambda \in (0,1])$. For the sake of simplicity and clear visualization, in our further considerations  we will present the limiting cases of Hartley entropy $\lambda =1$ and standard Shannon entropy ($\lambda=0$), as well as an intermediate value of $\lambda=1/2$ \cite{GazeauTsallis2019,Tsallis2020,Amigo2018,Duch2008,Wilk2000}. These entropies give a more detailed measure of correlations, and they are used to evaluate information in very different fields (for e.g. they are used for strongly correlated systems) \cite{tsallisbook}. Thus, for the Schwarzschild black hole, R\'enyi entropy results in
 \begin{figure}[h!]
\centering
\scalebox{0.39}
{\includegraphics{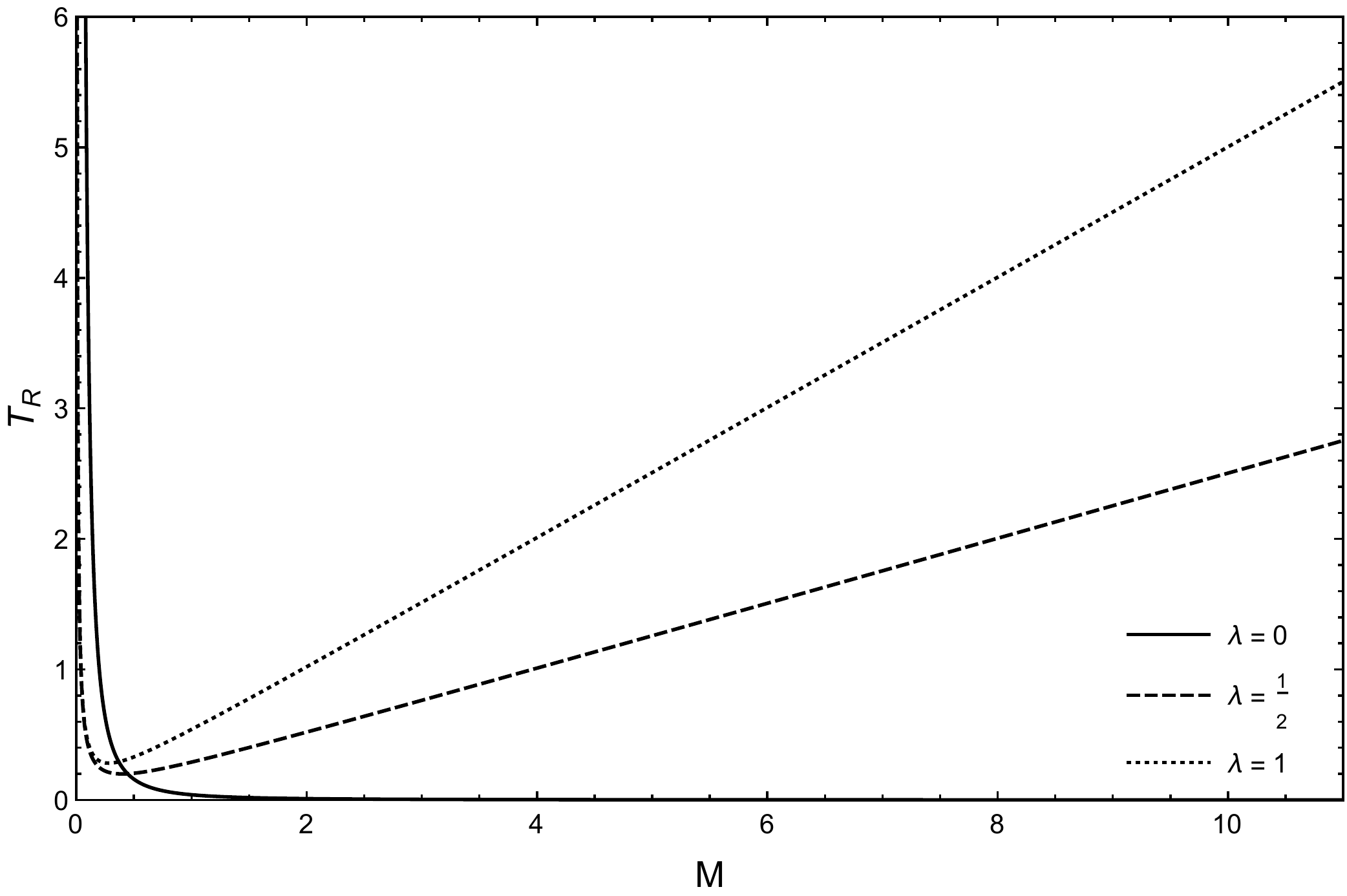}}
\caption{ R\'enyi temperature $T_R$ as a function of mass $M$ for different values of $\lambda=0,1/2,1.$  We have taken natural units such that $M_0=c=\hbar=k_B=G=m_p=1$.} 
\label{RTemp}
\end{figure}
\begin{figure}[h!]
	\centering
	\scalebox{0.39}
	{\includegraphics{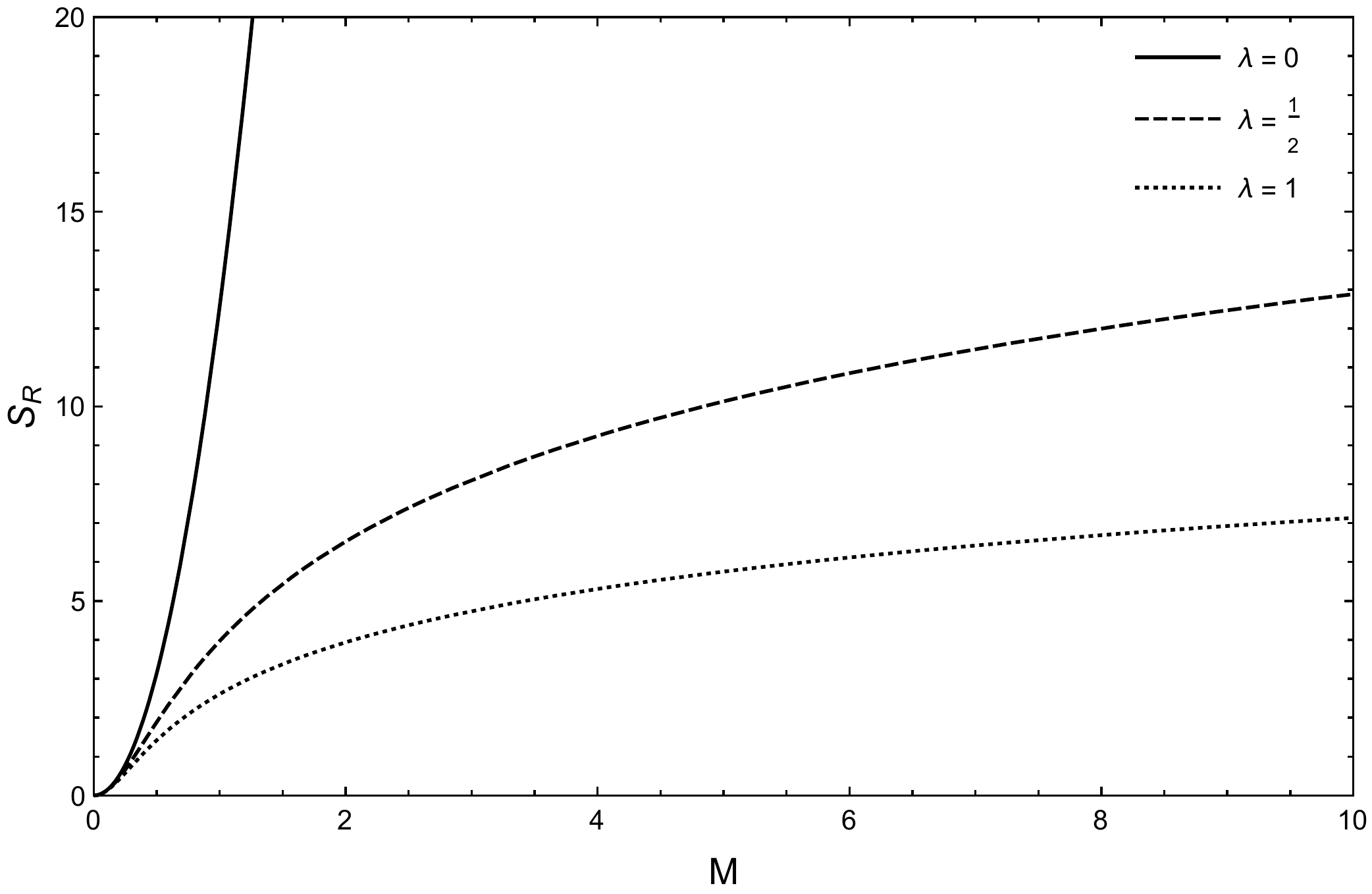}}
	\caption{R\'enyi entropy $S_R$ versus mass $M$ for different values of  $\lambda$. Bekenstein entropy corresponds to $\lambda=0$, and the studied two orders of R\'enyi entropy to $\lambda =1/2$ $(q=\frac{1}{2})$, and $\lambda =1$ $(q=0)$. We have taken natural units such that $M_0=c=\hbar=k_B=G=m_p=1$.}
	\label{REntropy}
\end{figure}
\be
S_R=\frac{k_B}{\lambda}\text{ln}\left[1+4\lambda \pi \left(\frac{M}{m_p}\right)^2\right]. 
\label{sr}
\ee
Note that the limit of Bekenstein entropy can be directly recovered by taking $\lambda \to 0$.
By using the first law of thermodynamics $c^2dM=T_RdS_R$, we can derive the corresponding R\'enyi temperature associated with the  Schwarzschild black hole, yielding
\be
T_R = T_{bh} + T_{\lambda}= \frac{c^3 \hbar}{8\pi k_B GM}+\frac{\lambda}{2}  \frac{Mc^2}{k_B}, 
\label{tr}
\ee
where the first term in the above equation is the Hawking temperature $T_{BH}$ and the second term, $T_\lambda$, comes as a consequence of the nonextensive nature of the R\'enyi entropy related to the introduction of the nonextensivity parameter $\lambda$. 

The evolution of the temperature (\ref{tr}) along the evaporation process for different values of parameter $\lambda$ is represented in Fig. \ref{RTemp}. It can be seen that R\'enyi temperatures are high for macroscopic masses, in contrast to Hawking temperature ($\lambda =0$), and that they decrease along with the evaporation. In opposition, they diverge at the last stages of evaporation in the same way as Hawking temperature, although for them it takes longer to start growing exponentially. These results indicate the thermodynamic features of the non-extensive R\'enyi entropy versus Bekenstein entropy. For R\'enyi entropy, it can be seen that consistently, the initial high temperature favors the radiation process. Note that the heat capacity associated with this R\'enyi temperature is positive \cite{stability1} so it is consistent with the decrease of temperature during evolution (In contrast to Hawking radiation). At the last stages of evaporation (for very small masses) this behavior changes and temperature and heat capacity coincide with the associated Hawking radiation, as it can be checked in the plot.

The evolution of entropy for different values of $\lambda$ can be seen in Fig. \ref{REntropy}. In this case, R\'enyi entropies ($\lambda \neq 0$) at initial states of evaporation present a lower value than Bekenstein entropy ($\lambda=0$). This fact can be understood as R\'enyi entropy considering interactions that are ignored in  Bekenstein-Hawking thermodynamical analysis. Then, the nonextensive thermodynamics shows much less uncertainty because of the measurement of correlations with this entropy, but their decrease is slower reaching the same evolution at the last stages of evaporation.

In order to introduce the phenomenological quantum gravity effects into R\'enyi entropy $S_R$, we consider the introduction of GUP in a similar way that enters into Bekenstein entropy. After some computations it results into the following expression for the GUP modified R\'enyi entropy, $S_{Rgup}$, 
\begin{widetext} 
\begin{equation}
S_{Rgup}=\frac{k_B}{\lambda} \ln{ \left\{ 1+ \pi \lambda\frac{ M^2}{ m^2_p } \left(2+\sqrt{4-\alpha_0\frac{ m_p^2}{M^2}}\right)  
\\
-  \pi \lambda \frac{\alpha_0}{2} \ln{ \left[ \frac{M}{M_0}  \left(2+\sqrt{4-\alpha_0\frac{ m_p^2}{M^2}}\right) \right] } \right\} }.
\label{srgup}
\end{equation}
\end{widetext}
The limits of R\'enyi entropy (when $\alpha_0 \to 0$) and Bekenstein entropy (when $\lambda \to 0$, that correspond, as assumed to Tsallis entropy) are recovered as expected.
By using $S_{Rgup}$, we can calculate the GUP modified R\'enyi temperature, $T_{Rgup}$, by using the relation, $c^2dM=T_{Rgup}dS_{Rgup}$, getting
\begin{widetext} 
\begin{equation}
T_{Rgup}= T_{gup} + T_{\lambda} \left\{ 1+ \alpha_0 \frac{\frac{m_p^2}{2M^2}}{2+\sqrt{4-\alpha_0\frac{ m_p^2}{M^2}}} \\
 \ln \left[\frac{M}{M_0} \left(2+\sqrt{4-\alpha_0\frac{ m_p^2}{M^2}}\right)\right] \right\} ,
\label{trgup1}
\end{equation}
\end{widetext}
that recovers the standard limits, as previously. It is worth to emphasize that GUP modifications to R\'enyi entropy and R\'enyi temperature predict the existence of a remnant in the same way as for the case of GUP modified entropy and temperature of Section \ref{GUP}.

\begin{figure}
	\centering
	\scalebox{0.40}
	{\includegraphics{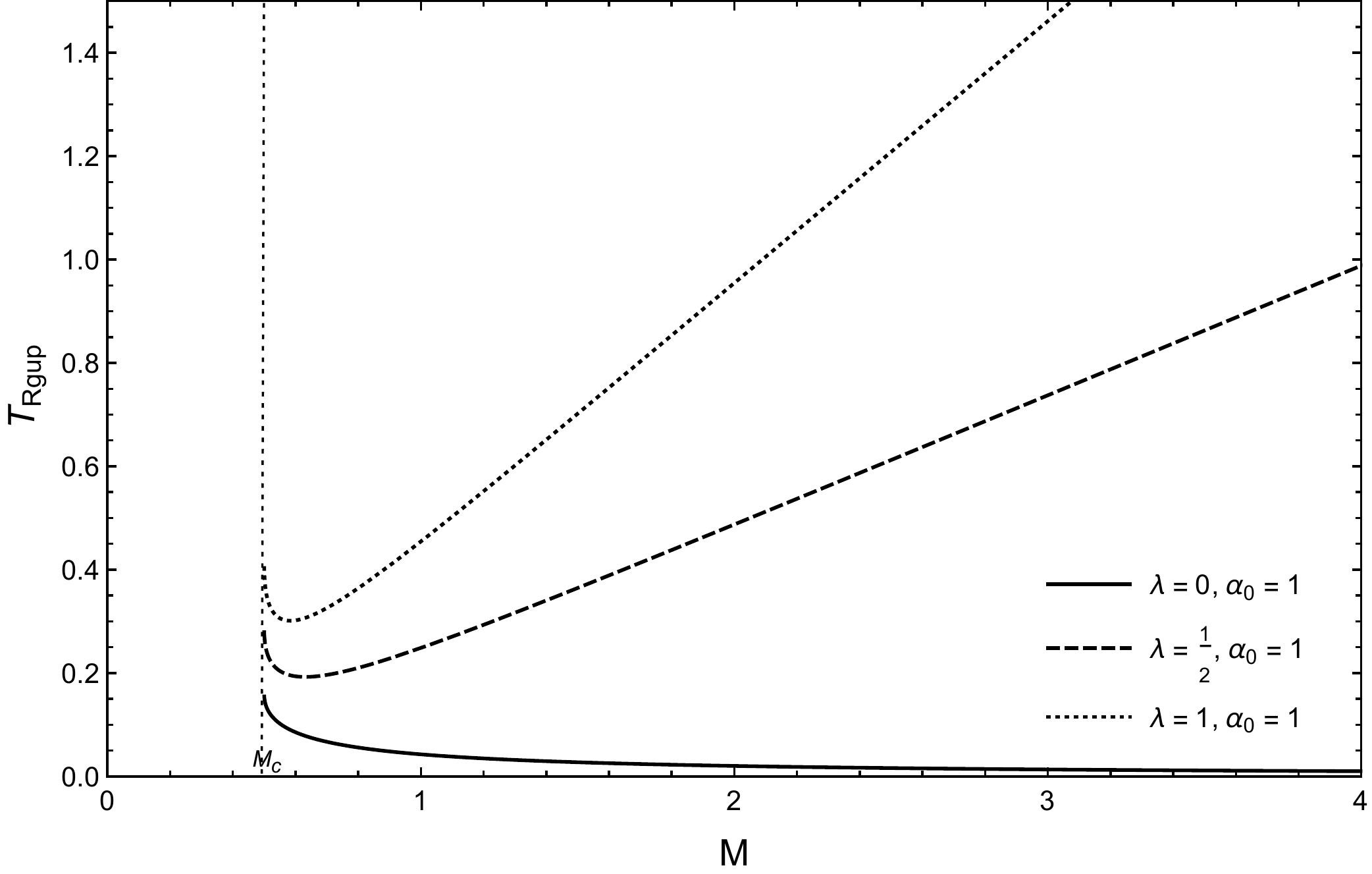}}
	\caption{GUP modified temperatures (with $\alpha_0=1$) as a function of mass $M$ for different values of $\lambda$ corresponding to GUP modified Hawking temperature $T_{gup}$ ($\alpha_0=1$ and $\lambda=0$) and GUP modified R\'enyi temperatures $T_{Rgup}$ ($\alpha_0= 1$ and $\lambda=1/2$, $1$).  We have taken natural units such that $M_0=c=\hbar=k_B=G=m_p=1$.} 
	\label{RTGUP}
\end{figure}

\begin{figure}
	\centering
	\scalebox{0.40}
	{\includegraphics{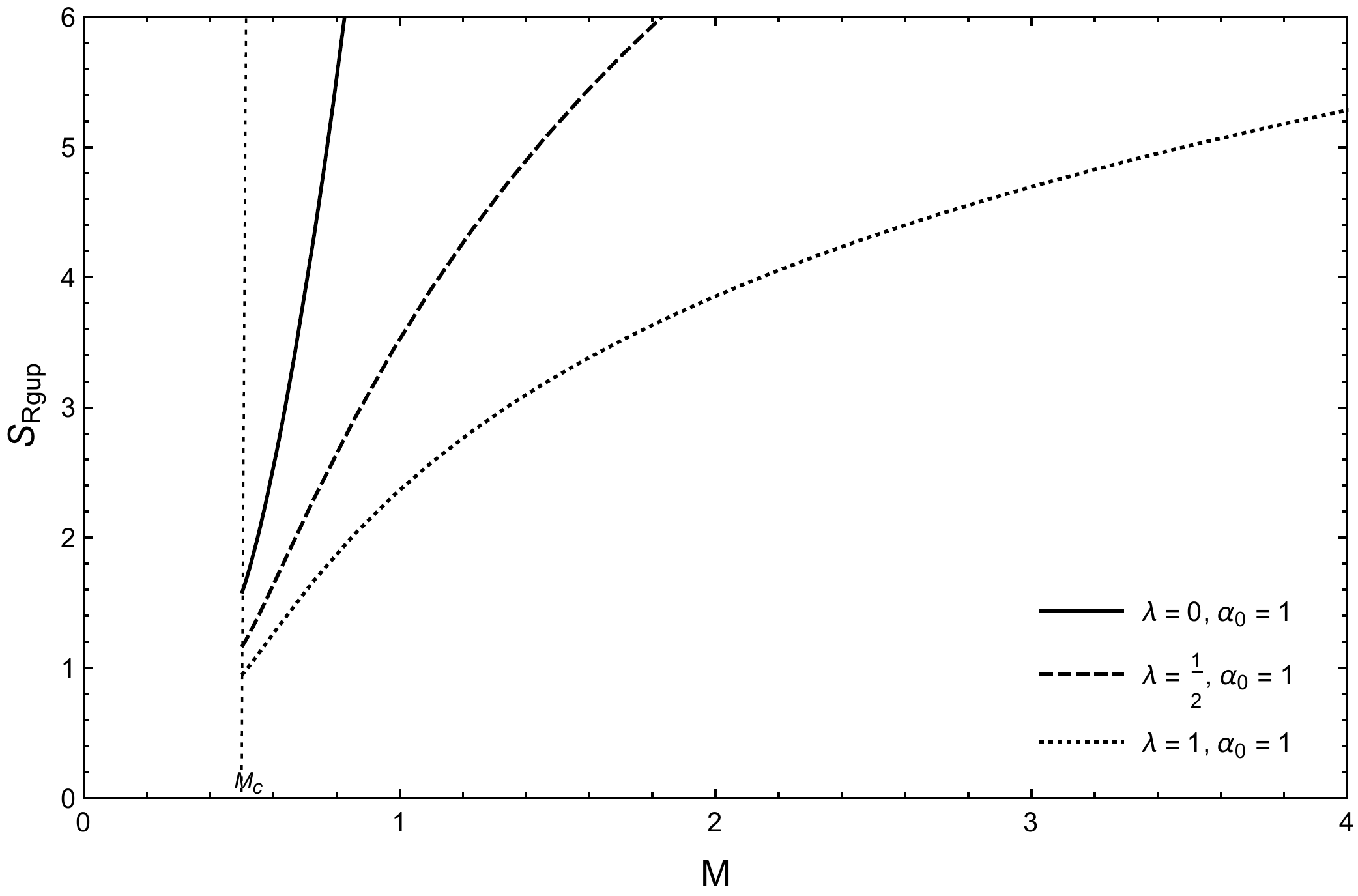}}
	\caption{GUP modified Bekenstein entropy $S_{gup}$  and GUP modified R\'enyi entropy $S_{Rgup}$ (with $\alpha_0=1$) versus mass $M$ for different values of $\lambda$. We have taken natural units such that $M_0=c=\hbar=k_B=G=m_p=1$.}
	\label{REGUP}
\end{figure}

In Fig. \ref{RTGUP} we show GUP modifications in temperature. It is direct to note that these quantum gravity effects modify all the temperatures in a similar way. Providing a final finite temperature for the remnant at $M_c$, with a value that is higher proportionally to parameter $\lambda$. The finite temperature of a remnant is fundamental to understand the final state and its thermodynamics. Note that a nonextensive remnant has a lower temperature.

The GUP modifications in the evolution of entropy are depicted in Fig. \ref{REGUP}. Due to the appearance of a remnant at $M_c$, the evaporation process finishes at that stage, corresponding in all the cases to a finite and small value of the entropy. It can be seen that the higher is $\lambda$, the lower is the entropy. These results show that there is less uncertainty on the state when analyzing evaporation using nonextensive parameters, then we can get more information from the evaporation that could help to shed some light on the information problem.

\section{ GUP Modified R\'enyi Entropy and the Sparsity of R\'enyi Radiation}
\label{sparsity}

\begin{figure}
\centering
\scalebox{0.40}
{\includegraphics{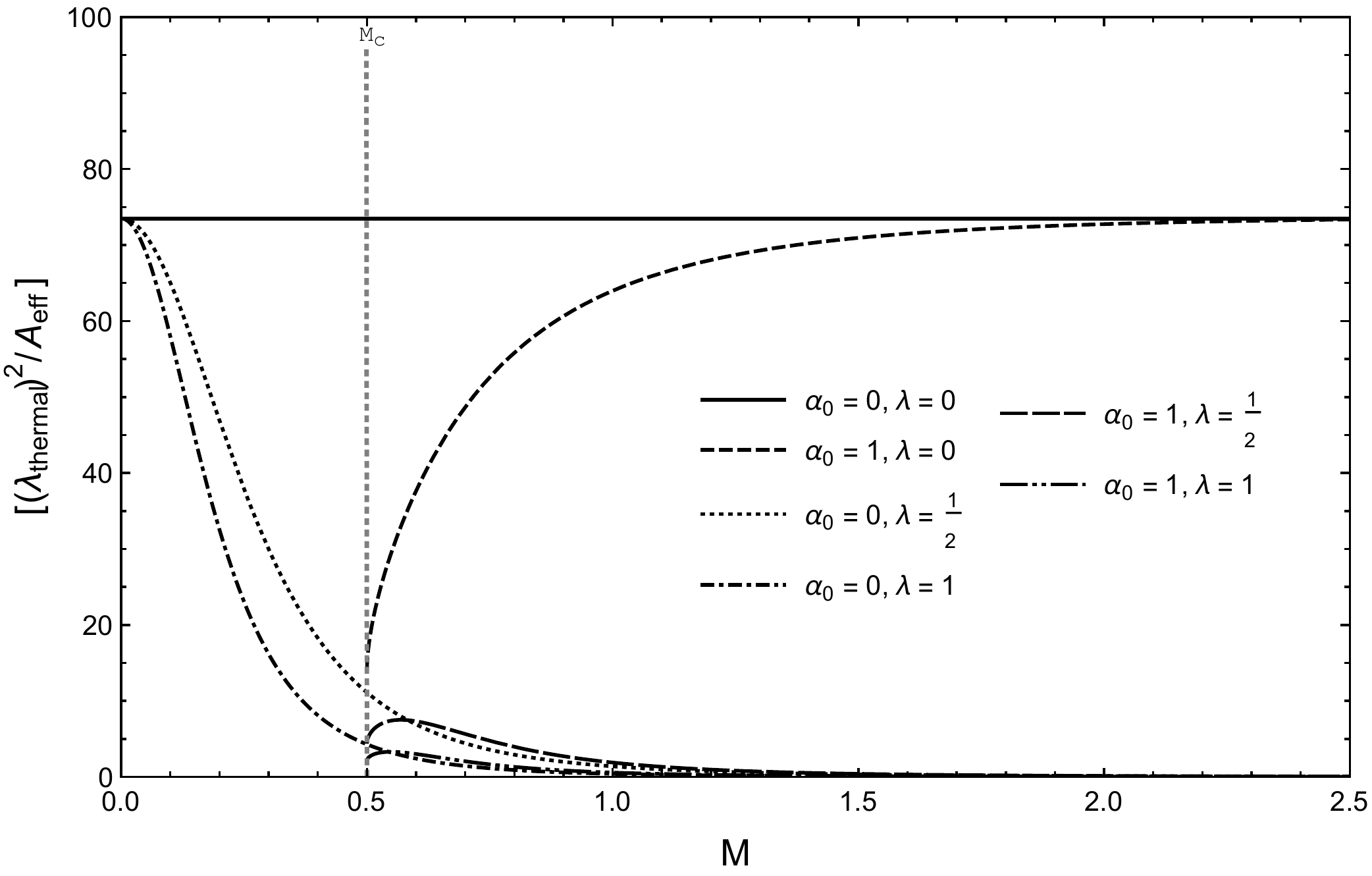}}
\caption{The sparsity parameter $\frac{\lambda^2_{thermal}}{A_{eff}}$ as a function of mass for different values of $\alpha_0$ and $\lambda$, corresponding to Hawking sparsity ($\alpha_0=\lambda=0$), GUP modified sparsity ($\alpha_0=1$ and $\lambda=0$), R\'enyi sparsity ($\alpha_0=0$ and $\lambda =0,1/2$) and GUP modified R\'enyi sparsity ($\alpha=1$ and $\lambda=0,1/2$), written in the legend respectively (from top to bottom). We have taken natural units such that $M_0=c=\hbar=k_B=G=m_p=1$. }
\label{RSparsity}
\end{figure}

It has been shown that Hawking radiation is very sparse throughout the whole Hawking evaporation process \cite{sparse1}. However, due to quantum gravity effects, it is no longer sparse at late stages of the black hole evaporation \cite{ana1,ana2}. That is, the sparsity decreases when the mass of a black hole approaches zero and quantum gravity effects are taken into account. The sparsity of the Hawking radiation is defined by a dimensionless set of parameters $\eta$ that in general are given by \cite{sparse1}
\begin{equation}
\eta=C\left[\frac{\lambda_{\text{thermal}}^2}{gA_{eff}}\right], 
\label{eta}
\end{equation}
where $C$ is a dimensionless constant that depends on the specific parameter $\eta$ chosen, $g$ is the spin degeneracy factor, $A_{eff}=\frac{27}{4}A$ is an effective area (that corresponds to the universal cross section at high frequencies), with $A$ the area of the black hole horizon, and the thermal wavelength, $\lambda_{\text{thermal}}$ reads
\be 
\lambda_{\text{thermal}}=2\pi\left[\frac{\hbar c}{k_BT}\right].
\label{lambdaT}
\ee
For the case of a Schwarszchild black hole one obtains
\be
\left[\frac{\lambda_{\text{thermal}}^2}{A_{eff}}\right]_{H}=\frac{64 \pi ^3}{27}>>1,
\ee
where the subscript $H$ refers to the consideration of Hawking temperature $T_{bh}$.  In order to calculate the sparsity, we express only the proportionality factor of thermal wavelength and effective area for the sake of generality. (For more details see \cite{sparse1}.) In this case, $\eta$ is much greater than one, showing a sparse radiation, in contrast to normal emitters in the laboratory.  In Fig. \ref{RSparsity} for $\alpha_0=\lambda=0$, the constant continuous line shows that the Hawking radiation is sparse throughout the black hole evaporation process. On the other hand, by incorporating the GUP modifications into $\eta$ \cite{ana1,ana2}, $T$ is replaced by $T_{gup}$ and $A_{eff}$ s replaced by
\be
A_{eff}=\frac{27}{4}A_{GUP}=\frac{27}{4}\left[A-\pi\alpha_0l_p^2\ln\frac{A}{A_0}\right], \label{Agup}
\ee
where $A_0= 16 \pi  G^2 M_0^2c^{-4}$ is an integration constant. Note that there is some discussion about the modifications that should be considered for the area \cite{ana1,onga}. In any case, one can develop the calculations in both ways to check that the difference in the final results would be only quantitative and not qualitative. Now we can write the GUP corrected sparsity parameter $\eta_{Hgup}$
\be
\left[\frac{\lambda_{\text{thermal}}^2}{A_{eff}}\right]_{Hgup}=\frac{64 \pi^3}{27}\times\frac{ M^2 \left[2+\sqrt{4-\alpha_0\frac{ m_p^2}{M^2}}\right]^2}{ \left[16 M^2-\alpha _0 m_p^2 \ln \left(\frac{M^2}{M_0^2}\right)\right]}. \label{etagup}
\ee
One can see that the sparsity depends on the mass of the black hole and that, at the initial stages of the black hole evaporation process, the Hawking flux is sparse but its sparsity decreases at the final stages of the evaporation (see Fig. \ref{RSparsity} for $ \alpha_0=1, \lambda=0$). It is worth mentioning that for negative values of the GUP parameter $\alpha_0$, the sparsity would increase during the final stages of the black hole evaporation process \cite{onga}.

In terms of R\'enyi temperature $T_R$, the sparsity parameter $\eta_R$ results in the expression
\be
\left[\frac{\lambda_{\text{thermal}}^2}{A_{eff}}\right]_{R}=\frac{64 \pi ^3}{27} \left[4\pi \lambda\frac{M^2}{m_p^2}+1\right]^{-2}. \label{etaR}
\ee
Then, introducing  GUP modifications to R\'enyi temperature $T_{Rgup}$, we derive $\eta_{Rgup}$ as
\begin{widetext}
\bea
\left[\frac{\lambda_{\text{thermal}}^2}{A_{eff}}\right]_{Rgup}&&=\frac{256 \pi ^3}{27 M^2 \left[16 M^2-\alpha _0 m_p^2 \ln \left(\frac{M^2}{M_0^2}\right)\right]}\nonumber\\ 
&&\times \left[\frac{\pi \lambda }{m_p^2}  \left\{2+\left(-2+\sqrt{4-\alpha_0\frac{ m_p^2}{M^2}}\right) \ln \left[\frac{M}{M_0} \left(2+\sqrt{4-\alpha_0\frac{ m_p^2}{M^2}}\right)\right]\right\}-
\frac{2}{\alpha_0 m_p^2}\left(-2+\sqrt{4-\alpha_0\frac{ m_p^2}{M^2}}\right)\right]^{-2}, \label{etaRgup}
\eea
\end{widetext}
 consistently recovering $\eta_{Hgup}$ for $\lambda \to 0$.
The sparsity behaviour of the radiation is completely different for R\'enyi temperatures. In this case, as it can be seen in Fig. \ref{RSparsity},  the R\'enyi radiation is not sparse at the intial stages of the evaporation process (so it is alike for normal emitters in the laboratory), but when reaching the last stages of evaporation, R\'enyi radiation starts being more and more sparse till the sparsity parameter associated with R\'enyi temperature reaches the same value as the sparsity parameter associated with Hawking temperature. This is consistent with previous analysis (remember that R\'enyi temperatures are much higher and heat capacity are positive for initial stages of evaporation). This behaviour at last stages change completely with the introduction of GUP modifications, as expected. Then, the effect is similar in all kind of radiations and GUP effects prevent radiation (in a general sense) from being sparse (being less sparse in proportion to $\lambda$). So, in the case of R\'enyi radiation, it would not be sparse at any moment of the evaporation, but would be emitted continuously. This result seems to indicate that when considering nonextensive thermodynamics and its quantum gravity correction at last stages of evaporation, one can see the black hole emitter is similar to a normal emitter during the whole evolution. That allows for a simpler and better understanding of the process and, moreover, it shows where distinctive features of black holes are codified, that is, in the nonextensive parameters and the underlying quantum gravity effects. This leads to the interest in further delving into the nonextensive thermodynamics of gravity in general and in particular of black holes.

\section{Conclusions}
\label{conclusion}
We have investigated the Generalized Uncertainty Principle impact on the R\'enyi entropy and temperature for the case of the Schwarzschild black hole. Furthermore, we have also studied the sparsity parameters of the radiation flux associated with R\'enyi and GUP modified R\'enyi temperatures.

 In fact, we have extended the issue of four fundamental constants $h, c, G, k_B$ as being simultaneously present in quantum thermostatistical expressions (\ref{tbh}) and (\ref{sbh}) as discussed in Ref. \cite{Tsallis2009} into an extra set of parameters: GUP parameter $\alpha_0$ and the nonextensivity of entropy parameter $\lambda$ (or $q = 1-\lambda$, cf. (\ref{ST})). The first of these two parameters adds a new important aspect of black hole physics aiming towards quantum gravity, while the second extends black hole thermodynamics onto the area, where non-extensive statistical mechanics applies (see Ref. \cite{reny4}).

 Our motivation was given by the fact that strong gravitational fields of black holes do not exhibit additivity of entropy and so the standard Boltzmann-Gibbs definition of entropy does not apply. Tsallis and R\'enyi entropies extend Boltzmann Gibbs statistics, so they may fit better the demand. 
 In the case of black holes, we assumed that Tsallis entropy is just the  Bekenstein entropy of black holes and that R\'enyi entropy could still further extend the range of application via the parameter $\lambda$ (or $q$) while having another nice property of fulfilling the generalized additive Abe rule (\ref{addL}). In such a construction, Tsallis entropy becomes the limit $\lambda \to 0$ of R\'enyi entropy which, on the other hand, gives the whole space of nonadditive property extensions of the black holes through the non-vanishing range of the nonextensivity parameter $\lambda$. 

We have shown that the black hole does not evaporate completely due to the minimum length modifications and hence, the associated GUP modified R\'enyi entropy and temperature have finite values at $M_c$ similar to the cases of GUP modifications to Bekenstein entropy and Hawking temperature. However, the behavior of both the temperatures and entropies are very different. At the initial stages, the specific heat capacity of black holes associated with nonextensive thermodynamics is positive, and for small masses, it becomes negative coinciding with the Hawking flux. This is reflected in higher initial temperature that will decrease during evaporation till the last stages of evaporation, where the semiclassical exponential growth of temperature is corrected by the introduction of quantum effects. Also, we see much lower initial entropy, that is, more information on the system, that decreased along with the evolution till a final finite value for the remnant.

Finally, we have analyzed that the radiation flux, corresponding to R\'enyi temperature, is not sparse at the initial stages of the black hole evaporation process, but the sparsity of R\'enyi radiation flux  increases during the black hole evaporation process and, in the end, it reaches the sparsity parameter of the Hawking flux. In addition to this, we have also shown that the modification of the sparsity parameter for the R\'enyi radiation flux due to the GUP corrected R\'enyi temperatures leads to the radiation flux that is less sparse than the Hawking flux.

Our analysis of nonextensive thermodynamics with quantum gravity effects results, globally, in a radiation flux that is not sparse in any moment of the evolution so it behaves similar to normal emitters, a temperature that decreases along with the evolution (with a positive heat capacity) and not very high entropy for black holes that slowly decreases till a finite value for the final remnant predicted by the theory.

In a more general framework, we should emphasize that the inclusion of quantum gravity aspects via $\alpha_0$ and nonextensivity aspects via $\lambda$ (or $q$) may have some more consequences onto the long-term discussion related to the problem of black hole information paradox resolution, through quantum black hole entanglement features widely discussed in the recent literature \cite{hr6,hr7,hr8,hr9,Gim2018,Hwang2018,Ana2018}. However, this matter is left for the scope of future papers.

\section*{Acknowledgements}

A. A-S. is funded by the Alexander von Humboldt Foundation. A.A-S work is also partially supported by the Project.  No.  MINECO FIS2017-86497-C2-2-P from Spain. M.P.D. wishes to thank Constantino Tsallis for discussions. This article is based upon work from COST Action CA15117 “Cosmology and Astrophysics Network for Theoretical Advances and Training Actions (CANTATA)”, supported by COST(European Cooperation in Science and Technology).


\begin{thebibliography}{99}

\bibitem{be} J.D. Bekenstein, Phys. Rev. D {\bf 7}, 2333 (1973).

\bibitem{hr1} S. W. Hawking,  Nature (London) {\bf 248}, 30 (1974).

\bibitem{hr2a}J. M. Bardeen, B. Carter, and S. W. Hawking, Commun. Math. Phys. {\bf31},  161 (1973)

\bibitem{hr2}S. W. Hawking, Commun. Math. Phys. {\bf 43}, 199 (1975).


\bibitem{hr3} S. W. Hawking, Phys. Rev. D {\bf 14}, 2460 (1976).

\bibitem{hr4} L. Susskind, L. Thorlacius, and J. Uglum, Phys. Rev. D {\bf 48}, 3743 (1993).

\bibitem{hr5}  G. t'Hooft, Nuc. Phys. {\bf B256}, 727 (1985). 

\bibitem{hr6} A. Almheiri, D. Marolf, J. Polchinski, and J. Sully, J. High Energy Phys. 02, (2013) 018.

\bibitem{hr7} A. Almheiri, D. Marolf, J. Polchinski, D. Stanford, and J. Sully, J. High Energy Phys. 09, (2013) 018.
 
\bibitem{hr8} P.  Chen,   Y.  C.  Ong,   D.  N.  Page,   M.  Sasaki,   and D. H. Yeom, Phys. Rev. Lett. {\bf 116}. 161304 (2016)

\bibitem{hr9} J. Maldacena and L. Susskind, Fortsch. Phys. {\bf 61}, 781 (2013).

\bibitem{hr10} S. W. Hawking, arXiv:1509.01147. 

\bibitem{hr10a} A. Almheiri, T. Hartman, J. Maldacena, E. Shaghoulian, and A. Tajdini, arXiv:2006.06872.

\bibitem{hr13} S. B. Giddings, Phys. Rev. D {\bf 46}, 1347 (1992).

\bibitem{hr14} J. Polchinski, in \emph{New Frontiers in Fields and Strings: Proceedings of the 2015 Theoretical Advanced Study Institute in Elementary Particle Physics}, edited by J. Polchinski, P. Vieira and O. DeWolfe (World Scientific, Singapore, 2017), p. 353.

\bibitem{hr15} P. Chen, Y.-Ch. Ong, D.H. Yeom,  Phys. Rep. {\bf 603}, 1 (2015)

\bibitem{hr16} A. Ejaz, H. Gohar, H. Lin, K. Saifullah, S. T. Yau,  Phys. Lett. B {\bf726}, 827-833 (2013)

\bibitem{hr11} W. G. Unruh and R. M. Wald, Rep. Prog. Phys. {\bf 80}, 092002 (2017).

\bibitem{hr12} D. Marolf, Rep. Prog. Phys. {\bf 80}, 092001 (2017).


\bibitem{gup1} R. J. Adler, P. Chen, and D. I. Santiago, Gen. Relativ. Gravit. {\bf 33}, 2101 (2001). 

\bibitem{GUPReview} S. Hossenfelder, Living Rev. Relativity {\bf 16}, 2 (2013). 

\bibitem{qg1} D. Amati, M. Ciafaloni, and G. Veneziano,  Phys. Lett. B {\bf 197}, 81 (1987).

\bibitem{qg2}David J. Gross and Paul F. Mende,  Phys. Lett. B {\bf 197}, 129 (1987).

\bibitem{qg3}  D. Amati, M. Ciafaloni, and G. Veneziano, Phys. Lett. B {\bf 216}, 41 (1989).

\bibitem{qg4} K. Konishi, G. Paffuti, and P. Provero, Phys. Lett. B {\bf 234}, 276 (1990).

\bibitem{qg5} A. Kempf, G. Mangano, and R. B. Mann, Phys. Rev. D {\bf 52}, 1108 (1995)
 
\bibitem{qg6} C. Rovelli, Phys. Rev. Lett. {\bf77}, 3288 (1996).

\bibitem{qg7} K. A. Meissner, Class. Quant. Grav. {\bf21}, 5245 (2004).

\bibitem{qg8} C. Bambi and F. R. Urban, Class. Quant. Grav. {\bf 25}, 095006 (2008).


\bibitem{gup2}  F.Scardigli,
Phys. Lett. B {\bf 452}, 39 (1999). 

\bibitem{gup3}  F.Scardigli and R.Casadio, 
Class. Quant. Grav. {\bf 20}, 3915 (2003).

\bibitem{gup4} F.Scardigli, G.Lambiase, and E.Vagenas 
Phys. Lett. B {\bf 767}, 242 (2017).


\bibitem{gup5} F. Scardigli, Class. Quantum Grav. {\bf 14}, 1781 (1997).  

\bibitem{gup6} F. Scardigli, P. Chen, and C.Gruber, Phys. Rev. D {\bf 83}, 063507 (2011).

\bibitem{gup7} F. Scardigli, Nuovo Cim. B {\bf 110}, 1029 (1995).

\bibitem{gup8} F. Scardigli, arXiv:0809.1832. 
 
\bibitem{correction1} G. Gour and A. J. M. Medved, Class. Quant. Grav. {\bf 20}, 3307 (2003).

\bibitem{correction2} A. J. M. Medved and E. C. Vagenas, Phys. Rev. D {\bf 70}, 124021 (2004).

\bibitem{sparse1} F.~Gray, S.~Schuster, A.~Vanabrunt, and M.~Visser, Classical Quantum Gravity  {\bf 33}, 115003 (2016).

\bibitem{sparse2} D. N. Page, Phys. Rev. D{\bf 13}, 198 (1976); {\it ibidem} D{\bf14 }, 3260 (1976); {\it ibidem} D{\bf16}, 2402 (1977).

\bibitem{sparse5}J. T. Firouzjaee and G. F. R. Ellis, Eur. Phys. J. C {\bf76}, 620 (2016).

\bibitem{ana1} A. Alonso-Serrano, M.P. D\c{a}browski, and H. Gohar, Phys. Rev. D{\bf 97}, 044029 (2018).

\bibitem{ana2} A. Alonso-Serrano, M.P. D\c{a}browski, and H. Gohar, Int. Journ. Mod. Phys. D{\bf 27}, 1847028 (2018). 


\bibitem{ong} Y. Ong, Phys. Lett. B{\bf 785}, 217 (2018).

\bibitem{greybody} A. Chowdhury and N. Banerjee, Phys. Lett. B{\bf 805}, 135417 (2020).

\bibitem{bibhas} A. Paul and B. R. Majhi, Int. J. Mod. Phys. A {\bf32}, 1750088 (2017). 

\bibitem{sparse6} S. Schuster, arXiv:1910.07256. 

\bibitem{sparse7} S. Schuster, Black Hole Evaporation: Sparsity in Analogue and General Relativistic Space-Times, Ph.D. thesis, arXiv:1901.05648. 

\bibitem{sparse8} Z. Feng, X. Zhou, S. Zhou and D. Feng, Annals Phys. {\bf 416}, 168144 (2020).

\bibitem{davies} P. C. W. Davies, Proc. R. Soc. Lond. A {\bf 353}, 499 (1977).

\bibitem{stability1} V.G. Czinner and H. Iguchi, Phys. Lett. B {\bf 752}, 306 (2016).

\bibitem{chinernew} V.G. Czinner and H. Iguchi, , Eur. Phys. J. C{\bf 77}, 892 (2017).

\bibitem{stability1a} G. Arcioni and E. Lozano-Tellechea, Phys. Rev. D {\bf 72}, 104021 (2005).

\bibitem {stability1b} O. Kaburaki, I. Okamoto, and J. Katz, Phys. Rev. D{\bf 47}, 2234 (1993).

\bibitem{biro} T.S. Biro and V.G. Czinner, Phys. Lett. B {\bf 726}, 861 (2013).

\bibitem{reny6} V.G. Czinner, Int. J. Mod. Phys. D{\bf 24}, 154 (2015). 


\bibitem{reny7} T.S. Biro and P. Van, Phys. Rev. E {\bf 83}, 061147 (2011).


\bibitem{reny4} C. Tsallis and L.J.L. Cirto, Eur. Phys. J. C{\bf 73}, 2487 (2013).


\bibitem{reny2} A. Renyi, Acta Math. Acad. Sci. Hung. {\bf 10}, 193 (1959). 

\bibitem{reny3} A. Renyi, {\it Probability Theory} (Elsevier: Amsterdam, The Netherlands, 1970).

\bibitem{reny1} C. Tsallis, J. Stat. Phys. {\bf 52}, 479 (1988).

\bibitem{tsallisbook} C. Tsallis, {\it Introduction to Non-Extensive Statistical Mechanics: Approaching a Complex World} (Springer, Berlin 2009).

\bibitem{Tsallis2009} C. Tsallis, Brazilian Journ. Phys. 39, 337 (2009). 

\bibitem{onga}Y. Ong, JHEP {\bf10}, 195 (2018).

\bibitem{ongb}Y. Ong, Phys. Rev. D {\bf98}, 126018 (2018).

\bibitem{ongc} Y. Ong, JCAP {\bf09}, 015 (2018).

\bibitem{alpha1} F. Scardigli and R. Casadio, Eur. Phys. J. C {\bf 75}, 425 (2015).

\bibitem{alpha2} D. Gao and M. Zhan, Phys. Rev. A {\bf 94}, 013607 (2016).

\bibitem{alpha3} Z.-W. Feng, S.-Z. Yang, H.-L. Li, and X.-T. Zu, Phys. Lett. B {\bf 768}, 81 (2017).


\bibitem{alpha4} P. Bosso, S. Das, and R.B. Mann, Phys. Lett. B {\bf 785}, 498 (2018).  

\bibitem{alpha5} D. Gao, J. Wang, and M. Zhan, Phys. Rev. A {\bf 95}, 042106 (2017). 

\bibitem{alpha6}S. Das and E.C. Vagenas, Phys. Rev. Lett. {\bf 101}, 221301 (2008).

\bibitem{alpha7} S. Giardino and V. Salzano, arXiv:2006.01580.

\bibitem{alpha8} G. Lambiase, F. Scardigli, Phys. Rev. D {\bf 97}, 075003 (2018).  

\bibitem{alpha9} S. Ghosh, Class. Quantum Grav. {\bf 31}, 025025 (2014).

\bibitem{reny8} S. Abe, Phys. Rev. E {\bf 63}, 061105 (2001).

\bibitem{GazeauTsallis2019} J.-P. Gazeau and C. Tsallis, Entropy {\bf 21}, 1155 (2019). 

\bibitem{Tsallis2020} C. Tsallis, Entropy {\bf 22}, 17 (2020).

\bibitem{Amigo2018} J.M. Amigo, S. G. Balogh, and S. Hernandez, Entropy {\bf 20}, 813 (2018). 

\bibitem{Duch2008} T. Maszczyk and W. Duch, Lecture Notes in Computer Science, {\bf 5097}, 643 (2008). 

\bibitem{Wilk2000} G. Wilk and Z. W{\l }odarczyk, Phys. Rev. Lett. {\bf 84}, 2770 (2000). 

\bibitem{Gim2018} Y. Gim, H. Um, and W. Kim, Journal Cosm. Astrop. Phys. {\bf 02}, 060 (2018). 

\bibitem{Hwang2018} J. Hwang, H. Park, D.-H.Yeom, and H. Zoe, J. Korean Phys. Soc. {\bf 73}, 1420 (2018). 

\bibitem{Ana2018} A. Alonso-Serrano and M. Visser, Phys. Lett. B{\bf 776}, 10 (2018). 








\end{thebibliography}
\end{document}